\documentclass[runningheads]{llncs}
\usepackage{graphicx}

\usepackage{listings}

\usepackage{xcolor}
\begin{document}
\title{System measurement of Intel AEP Optane DIMM }

\author{Tianyue Lu\inst{1} \and
	Haiyang Pan\inst{1} \and
	Mingyu Chen\inst{1}}

\institute{
	Institute of Computing Technology, Chinese Academy of Sciences, Beijing, China\\
	\email{\{lutianyue,panhaiyang,cmy\}@ict.ac.cn}}

\maketitle              
\begin{abstract}
In recent years, memory wall has been a great performance bottleneck of computer system. To overcome it, Non-Volatile Main Memory (NVMM) technology has been discussed widely to provide a much larger main memory capacity. Last year, Intel released AEP Optane DIMM, which provides hundreds of GB capacity as a promising replacement of traditional DRAM memory. But as most key parameters of AEP is not open to users, there is a need to get to know them because they will guide a direction of further NVMM research. In this paper, we focus on measuring performance and architecture features of AEP DIMM. Together, we explore the design of DRAM cache which is an important part of DRAM-AEP hybrid memory system. As a result, we estimate the write latency of AEP DIMM which has not been measured accurately. And, we discover the current design parameters of DRAM cache, such as tag organization, cache associativity and set index mapping. All of these features are first published on academic paper which are greatly helpful to future NVMM optimizations.

\end{abstract}
\section{Introduction}

In modern era, memory wall \cite{memory_wall} has been a great performance bottleneck to many big-data programs. Plenty of data is handled in parallel which requires great memory bandwidth and memory capacity. Multi-channel memory system has been a mainstream solution to provide both enough bandwidth and capacity. But to growing big-data applications, memory channels are not able to provide enough scalability. In particular, memory capacity is not well scalable in traditional DRAM memory system. This is due to that, DRAM capacity per-DIMM has reached 64GB and is hard to grow further \cite{twin-load}. New memory technology is demanded to break the limit of DRAM memory. 

In this case, Non-Volatile Memory (NVM) \cite{nvm} has been a research hotspot to solve memory capacity wall. NVM, includes PCM \cite{pcm}, RRAM \cite{rram}, MRAM \cite{mram}, provides much higher storage density than DRAM. Besides, some NVM medium provides similar read latency and bandwidth with DRAM, which means NVM could be a potential replacement of DRAM to compose memory system. NVDIMM \cite{nvdimm} is an JEDEC \cite{jedec} standard of using NVM as storage module. Among several branches, NVDIMM-P is a DRAM-NVM hybrid memory standard. In NVDIMM-P \cite{nvdimm-p}, DRAM and NVM are both deployed as memory channel when NVM is used as main memory while DRAM is used as cache. NVM has larger capacity and DRAM has better performance, so DRAM cache improves access performance of partial hot data and NVM meets the demand of memory capacity of applications. 

Intel AEP Optane \cite{aep} is a commercial product which achieves NVDIMM-P architecture. AEP DIMM is inserted on DIMM slot as main memory and DRAM DIMM on other slots are used as transparent off-chip cache. Each AEP DIMM has 128GB to 512GB capacity (depends on different product models ) which is much larger than DRAM. At the same time, DRAM cache still has 1/8 to 1/2 capacity of NVM main memory which means it can easily achieve a high cache hit rate. So with this memory system, programs which do not require large memory capacity still acquire same performance with the case of traditional pure-DRAM memory. And on the other side, programs which requires hundreds of GB memory could gain great performance improvement to the case that main memory is lacked and disk is used as swap memory. Unfortunately, as a commercial product, many performance parameters of AEP are not published. But in order to use AEP wisely, we need to know parameters fully, including r/w latency, bandwidth, and structure design parameters like DRAM cache associativity, cache tag placement and AEP access granularity. With detailed parameter, academia and industries can optimize the memory architecture and scheduling more carefully. 

Some prior works had made great work on AEP parameter measurement. \cite{memsys-optane} gives a detailed AEP access bandwidth and read latency. \cite{fast20-yang} shows that AEP has a write buffer on AEP DIMM which improves AEP write performance. But in most works, write latency is not well surveyed because write instructions are committed as soon as write data are written on CPU cache or committed on write command queue. Also, most work did not take into account DRAM cache architecture parameters. In many researches, DRAM cache optimization has been discussed \cite{dramcache1,dramcache2,dramcache3,tdvcache}, including cache associativity design, cache tag placement, cache replacement algorithm and so on. Different cache designs determine cache performance, and in real system, DRAM cache parameters are determined by memory controller on DRAM channel which is not open to most users. DRAM cache architecture is important and some of its parameters are able to be measured.

In this paper, we measure AEP and DRAM cache parameters. Different with prior works, we focus on the architecture parameter of DRAM-AEP hybrid memory system. But first, we estimate the write latency of AEP DIMM under different high write bandwidth. High write bandwidth makes all write command queues in every levels of memory hierarchy are full, so that new write instruction will not be able to issue in CPU instruction queue (actually is Re-Order Buffer). At this time, the measured latency on both ends of one write instruction is the time of write queues get one empty entry, which is nearly the completion time of one write command finishing data writing on AEP module. Furthermore, we measure the architecture parameter of AEP DIMM and DRAM cache. To AEP DIMM, we test if a small buffer is on DIMM module. And to DRAM cache, we measure the cache associativity and address mapping of cache set index. As a result, we find that a 16KB buffer is on AEP DIMM and it is fully-associative. And DRAM cache is direct-mapped cache whose tag and data is stored in one 64B cacheline and its set index mapping is also found. Detailed measurement results are introduced later in this paper.

In following paper, Section 2 introduces Intel AEP and its two deployment modes. Section 3 introduces our experiment platform and our test benchmarks. And measurement results are introduced in Section 4. Based on these results, Section 5 provide some discussions about DRAM-AEP hybrid memory systems, and Section 6 makes conclusions.

\section{Background}
In this section, we will introduce Intel AEP Optane DIMM in detail. 

\subsection{Intel AEP hardware architecture}
Figure 1 shows the architecture of Intel DRAM-AEP hybrid memory system. Both the AEP and DRAM DIMMs are accessed via traditional DDR-T buses and managed by memory controller hardware named iMC. All DRAM DIMMs are standard DDR4 DIMMs. As illustrated in [fast20], a on-DIMM controller named XPController manages the requests of Optane DIMM. XPController translates the request address into the real address on AEP DIMM and maintains a buffer named XPbuffer to accelerate access, similar to the function of row buffer on DRAM DIMMs. iMC manages the DIMMs in two modes including memory mode and App direct, and two modes can be set via ipmctl command in command line of Linux dynamically. 

\begin{figure}
	\centering
	\includegraphics[scale=0.9]{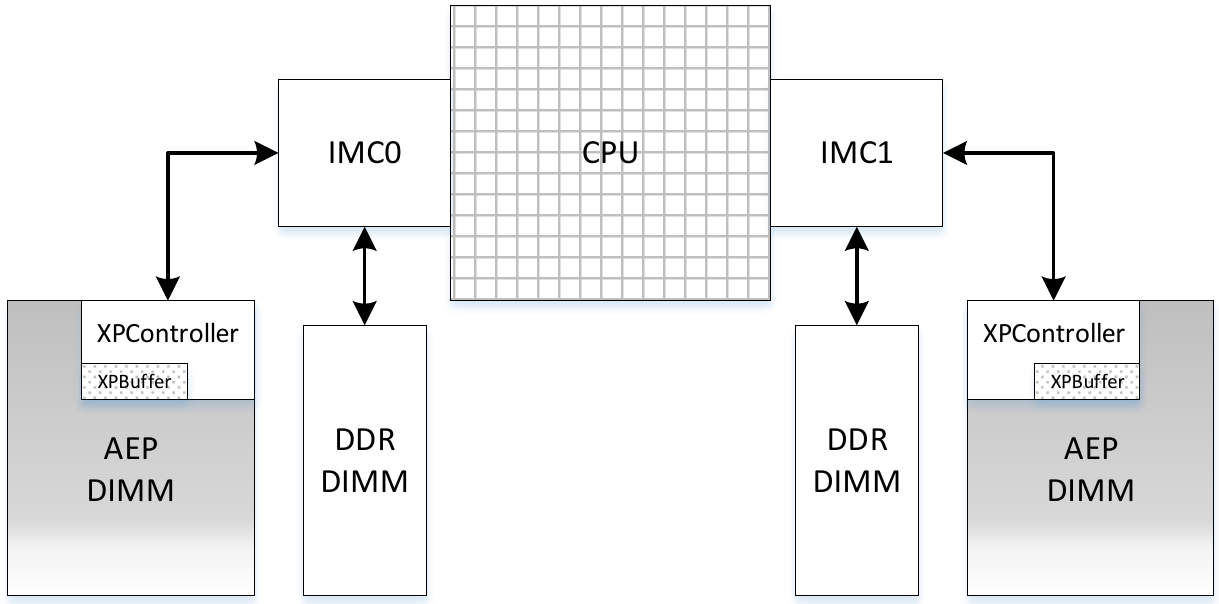}
	\caption{Hardware architecture of Intel AEP.}
	\label{structure_aep}
\end{figure}

\subsection{Two modes of AEP DIMMs}
Figure 2 shows the storage topology in different modes of AEP DIMMs. We can find that DRAM DIMMs also change their roles according to AEP modes. In memory mode, AEP DIMMs work as main memory and the DRAM DIMMs are cache of the AEP DIMMs on the same channel. Main memory capacity seen in OS is all capacities of AEP DIMMs while DRAM cache is transparent to software. 

In App direct mode, AEP DIMMs work as persistent device which can be direct accessed by applications, similar as a common Optane SSD device. In this case, AEP DIMMs are accessed as traditional block devices. Besides, Intel supports a library that can implements direct load/store instructions for AEP DIMMs using the fsdax file system. Persistency is ensured by Intel ADR (Asynchronous DRAM Refresh) mechanism, which ensures writing data in ADR domain will survive a power failure. In this architecture, ADR domain includes IMC and AEP DIMMs. To programers, adding clwb and sfence instruction behind write instruction will ensure writing data persistent, when clwb writes data from CPU cache to IMC and sfence insures clwb to be committed. In AEP App Direct mode, DRAM DIMMs act as the traditional volatile main memory and can be seen as main memory capacity in software. 

Public information and prior work have some introductions about both memory mode and App Direct mode, but there is no clear description or measurement on specific details. For example, how is the DRAM cache tag managed? What is the cache associativity and address mapping of set-index? What is the write latency from IMC to AEP module? We find these architecture parameters important, and we manage to measure them in this paper. Our methodology is introduced in next section.

\begin{figure}
	\centering
	\includegraphics[scale=0.9]{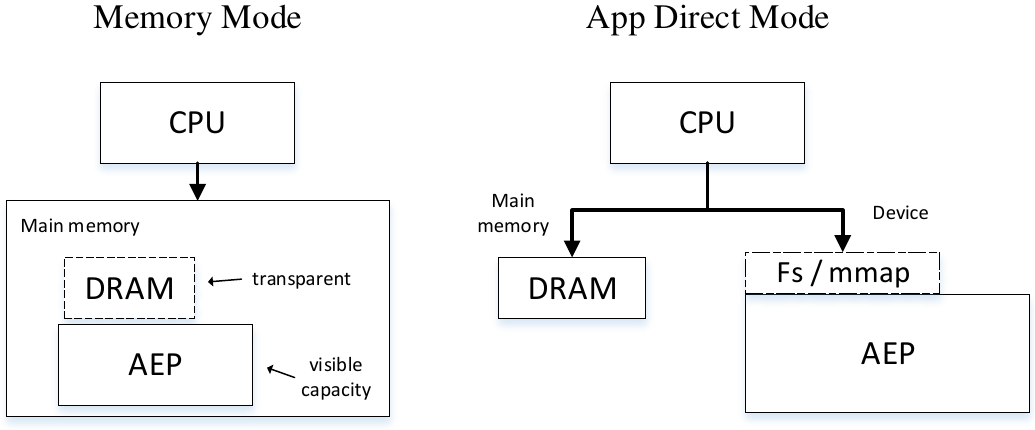}	
	\caption{Two modes of AEP.}
	\label{twomode}
\end{figure}

\subsection{DRAM Cache}
As mentioned above, DRAM is used as DRAM cache in AEP memory mode. Many prior researches have focused on DRAM cache design and optimization. To DRAM cache, there are two major problems to solve: hit rate and hit latency. On hit rate, it is not hard for DRAM cache to reach a hit rate above 50\%, as it has a capacity of 1/8 to 1/2 main memory. But on the other hand, it is hard to increase a hit rate that is already so high. And this hit rate is more critical to DRAM-AEP hybrid memory system performance than traditional one as the latency gap between DRAM cache and AEP main memory could be larger than it between CPU cache and DRAM main memory. On hit latency, DRAM cache must handle cache tags carefully because extra tag or metadata access issues extra DRAM access which is much more expensive than on-chip cache access. Considering hit rate and hit latency, replacement algorithm design on DRAM cache faces new challenges that, complex algorithm like LRU or RRIP maintains a large amount of metadata which brings large storage overhead and access overhead. Simple algorithm like random or FIFO algorithm could not achieve a high hit rate. So in our paper, we are concerned about the actual DRAM cache design with Intel iMC which we think will provide a direction of further DRAM cache research. 

\section{Experimental platform}
In this section, we will introduce our measurement platform and benchmark. 

\subsection{Platform configuration}
In our experiment, we deploy Intel server with 1-channel DRAM and 1-channel AEP. Detailed server configuration is shown in Table \ref{configuration}. DRAM has 16GB capacity and AEP has 128GB. Both DIMMs are inserted on slots managed by same IMC. And our benchmarks are bound on cores of local CPU. So in our experiment, we avoid the impact of NUMA architecture to get the local AEP performance parameters. 

\begin{table}[htbp]
	\caption{Experiment Platform Configurations}
	\label{configuration}
	\centering
	\begin{tabular}{l|l}
		\hline
		Processor & Intel(R) Xeon(R) Gold 6246 CPU @ 3.30GHz \\
		\hline
		Operating System & Linux Kernel 4.20 in CentOS 7\\
		\hline
		LLC Cache Size & 24.75MB \\
		\hline
		DRAM & 16GB per DIMM \\
		\hline
		NVM & 128GB per DIMM \\
		\hline
	\end{tabular}
\end{table}

\subsection{Software benchmark}
In measurement programs, we use RDTSC instruction between two ends of every read/write instruction to measure latency. Instead of average value, we get a variation curve of latency value on the timeline. As a result, some latencies show a stable value, but on the other side, some values are varied regularly.

For read latency measurement, we use single-thread program to access malloc memory space of main memory in AEP memory mode of mmap space in App Direct mode. All accessed memory addresses are chained. For example, We first read address A, and data in A is B, which is next access address. In this case, access B must wait for access A to be finished. So there will be dependency among all reads to avoid read latency to be hidden in pipeline. To make read instruction completed, LFENCE is added before RDTSC instruction. The following code shows an example of our testing.

\lstset{language=C}
\begin{lstlisting}
unsigned long test()
{
    unsigned long timehi, timelo;
    unsigned long start;
    asm volatile ("lfence\n");
    asm volatile ("rdtsc":"=a"(timelo), 
          "=d"(timehi):);
    start = timehi << 32 | timelo;
    asm volatile ("lfence\n");
     
    code_for_testing;
    
    unsigned long end;
    asm volatile ("lfence\n");
    asm volatile ("rdtsc":"=a"(timelo),
          "=d"(timehi):);
    start = timehi << 32 | timelo;
    asm volatile ("lfence\n");
    
    return end - start
}
\end{lstlisting}

To measure write latency, we need to ensure that write bandwidth is high enough to fill all memory write queues. In this case, new write instruction is not allowed to issue as there is no space for new writes. Only at this time, the completion time of a write instruction reflect the write latency on AEP module. All instructions are issued and committed in Re-Order Buffer (ROB). And, write instruction is committed at the time of that it has been sent on memory queue. Only when queue is full, the write instruction will wait in ROB. The waiting time depends on how much time it takes for a queue entry is deleted. And only when writing data is written on AEP module, the memory queue on IMC is allowed to delete corresponding write command. Therefore, we use three threads to fill memory bandwidth and memory queues. And fourth thread measures the completion time of a write instruction. Three threads are enough to fill memory bandwidth, which is confirmed in our AEP bandwidth experiment in Section 4.2. 

\subsection{Lecroy Kibra DDRSuite Tool}
To deal with DRAM cache measurement, as it is transparent device to OS, we use Lecroy Kibra DDR Protocol Analyzer Suite to support our experiment. Lecroy Kibra DDR Protocol Analyzer Suite is connected with server board via standard DIMM slot, and DRAM DIMM is inserted on Lecroy Analyzer Suite as shown in Figure X. At work, DRAM DIMM still work as normal memory module, and DDR command sent to it can be caught by Lecroy Analyzer Suite. Command type (Read/write/activate/precharge) and access address can be get, excluding R/W data. In AEP memory mode, DRAM acts as cache. And we use Lecroy Analyzer Suite to catch the DDR commands from IMC, to observe the pattern of it. Analyzing the pattern will help us know the workflow of DRAM cache, like how much DDR commands are needed during a cache hit/miss.

Besides, to test DRAM cache set index mapping, we use a pair memory addresses, which have only one-bit difference. Sometimes when some address bits flip, two addresses conflict in cache set, and sometimes they don’t. Based on this experiment, we can find out how memory addresses are mapped to DRAM cache set index. 

\begin{figure}
	\centering
	\includegraphics[scale=0.35]{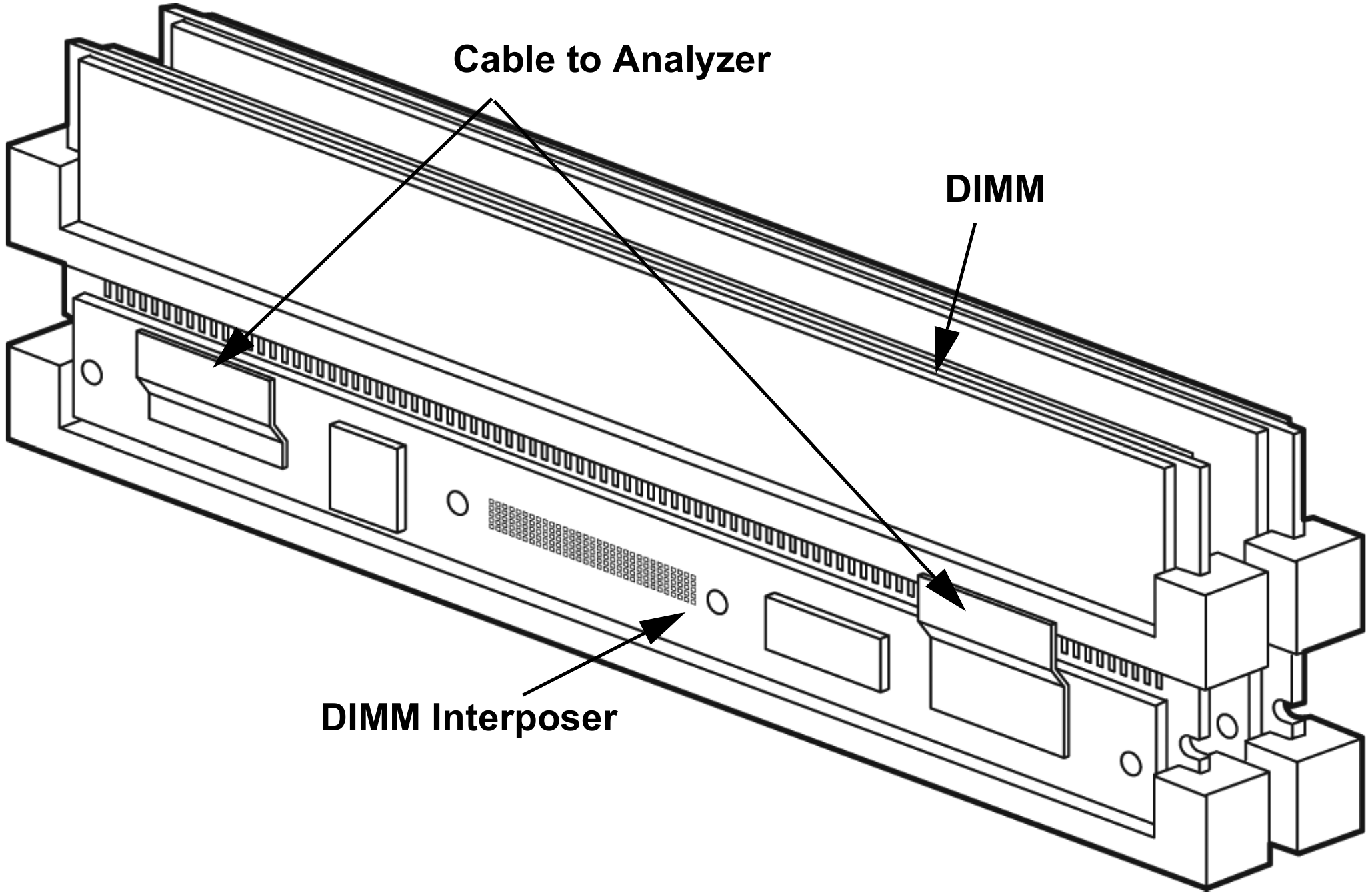}	
	\caption{Two modes of AEP.}
	\label{twomode}
\end{figure}

\section{Experiment results}
Our results will be shown in this section. First, we show measured AEP read/write latency in two modes. And based on latency results, we analyze the results and obtain some architecture features of AEP DIMM. Also, on memory mode, we measure how DRAM cache is organized including cache data and tag. At last, limit bandwidths of AEP are educed. 

\subsection{Read latency of AEP and XPbuffer feature}
We measure AEP latency in different access modes. Both read and write are measured, and sequential accesses or random accesses also give different results.

\begin{figure}
	\centering
	\includegraphics[scale=0.5]{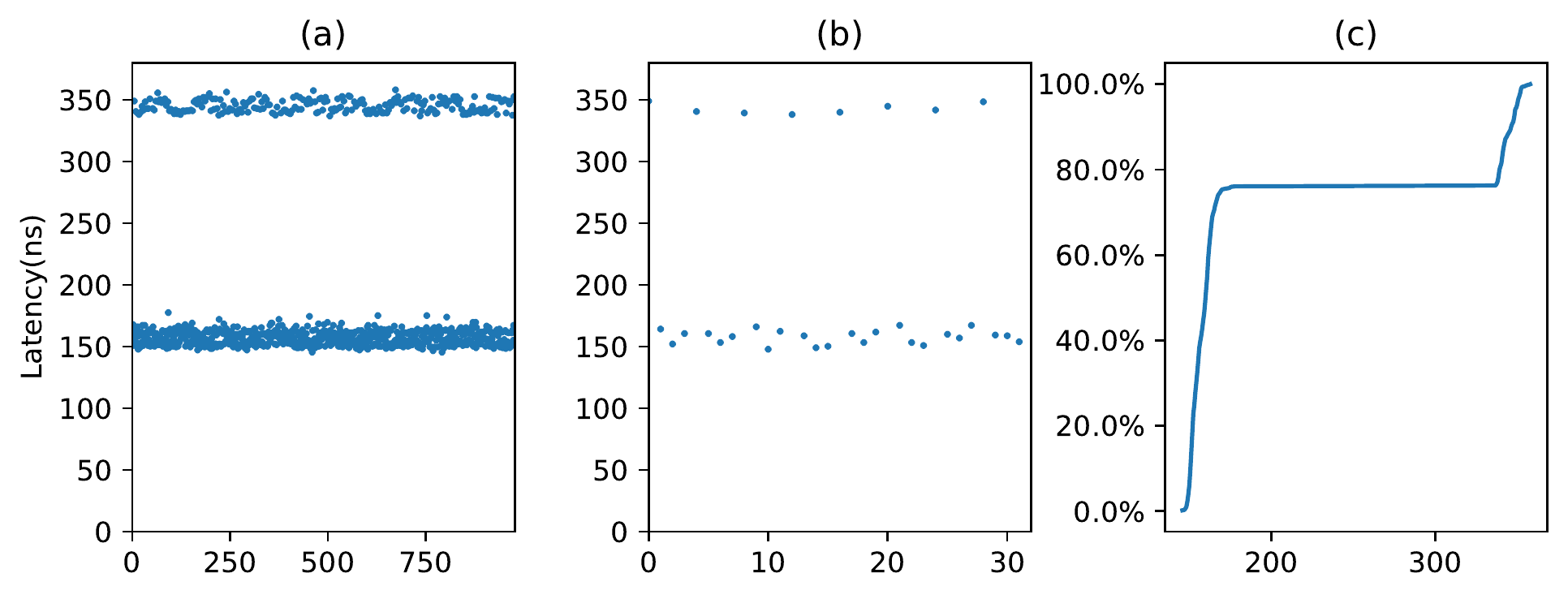}	
	\caption{Sequential Read Latency in APP direct mode.}
	\label{appdirect_seq_read_latency}
\end{figure}

First, we measure read latency in AEP App Direct mode. Figure \ref{appdirect_seq_read_latency} shows the results of sequential read. In Figure \ref{appdirect_seq_read_latency}(a), we find that read latency has two alternate values, the lower one is about 150ns, and the higher one 350ns. We take a consecutive 32 points in Figure \ref{appdirect_seq_read_latency}(a) and generate Figure \ref{appdirect_seq_read_latency}(b). Actually, we find that read latency values have a 3-low with 1-high pattern. As shown in Figure \ref{appdirect_seq_read_latency}(c), lower value takes 75\% and other 25\% is higher value. We think that the lower value means the read accesses hit on XPbuffer of AEP DIMM. 

\begin{figure}
	\centering
	\includegraphics[scale=0.5]{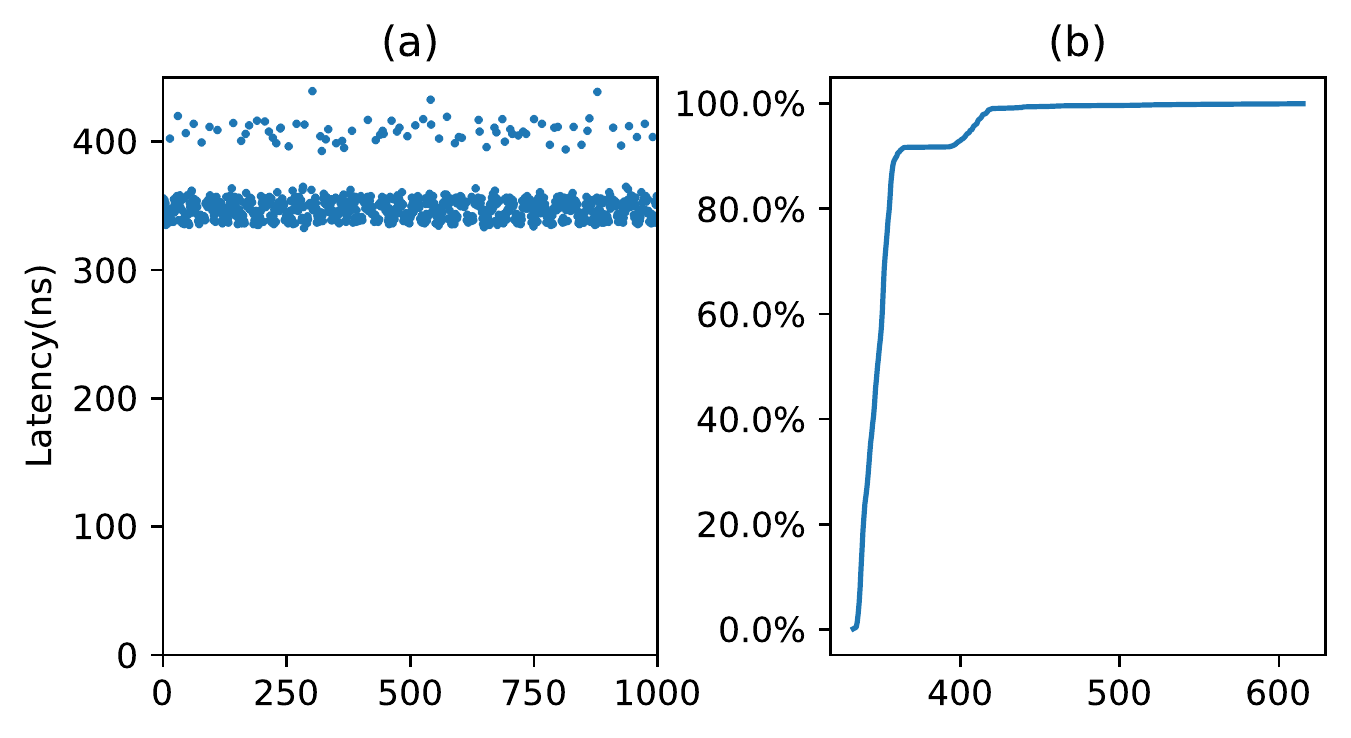}	
	\caption{Random Read Latency in APP direct mode.}
	\label{appdirect_rand_read_latency}
\end{figure}

\begin{figure}
	\centering
	\includegraphics[scale=0.5]{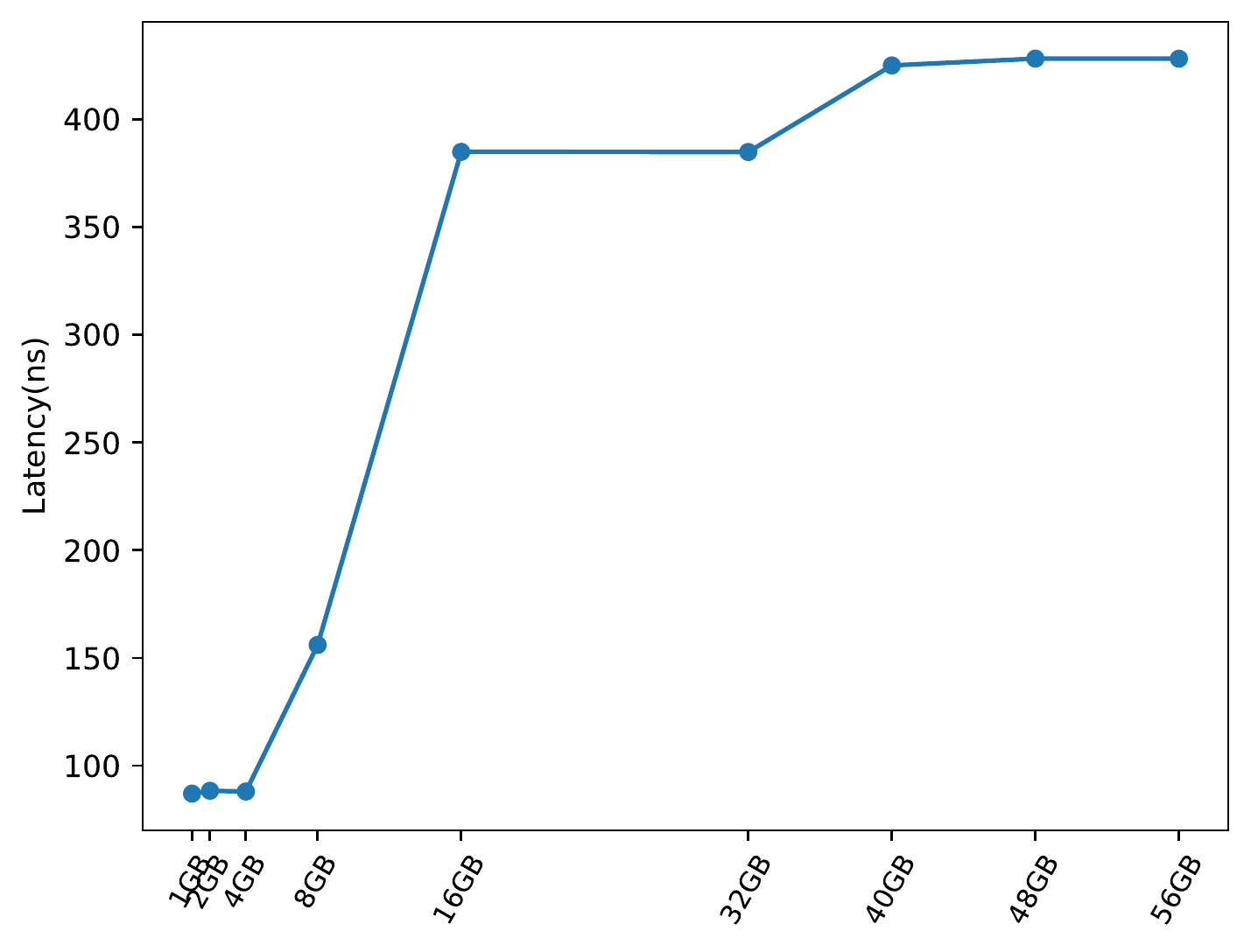}	
	\caption{Random Read Latency in Memory Mode over different footprints.}
	\label{memory_rand_read_latency}
\end{figure}

The reason is, each time the latency value shows a high value at about 350ns, the access address of sequential reads comes to 256-aligned one. And it hints us that, the access granularity of AEP is 256B. This means that, in AEP module, each DDR read commands get 256B data from module. Although only 64B is returned through DDR bus, other 192B is buffered on XPbuffer AEP DIMM.

Random reads show a stable latency value, as shown in Figure \ref{appdirect_rand_read_latency}. Different with sequential reads, latency values are much more stable, and equal to the higher value in the results of sequential reads. In summary, we think real read latency of AEP DIMM is about 350ns. 

In AEP memory mode, we get almost the same latency results with App Direct mode. Besides, we run our test benchmark with different memory footprints. And with small footprints, all accesses will hit on DRAM cache in memory mode. Figure \ref{memory_rand_read_latency} shows the measured latencies with different footprints. Leads to a value jumping happens at 16GB, the memory footprint is larger than DRAM cache capacity which causes all accesses are miss on DRAM cache. Lower than 16GB, the latency value equals to DRAM read latency, which is about 90ns.

\subsection{Write Latency of AEP}
\begin{figure}
	\centering
	\includegraphics[scale=0.7]{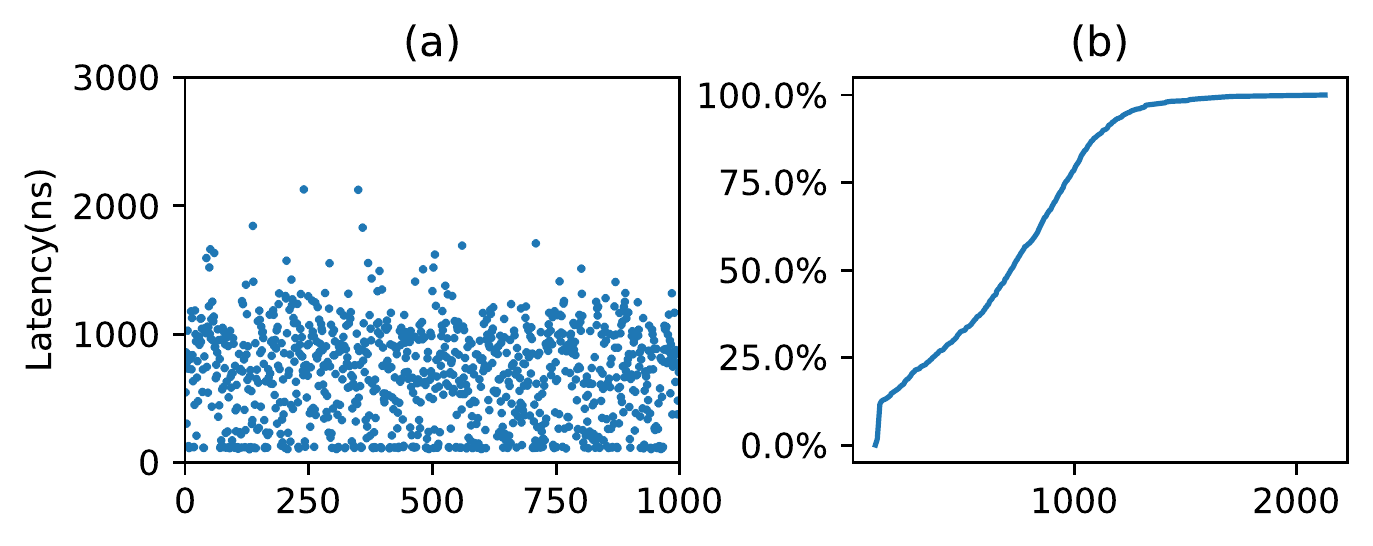}	
	\caption{Sequential Write Latency in APP direct mode.}
	\label{appdirect_seq_write_latency}
\end{figure}

\begin{figure}
	\centering
	\includegraphics[scale=0.7]{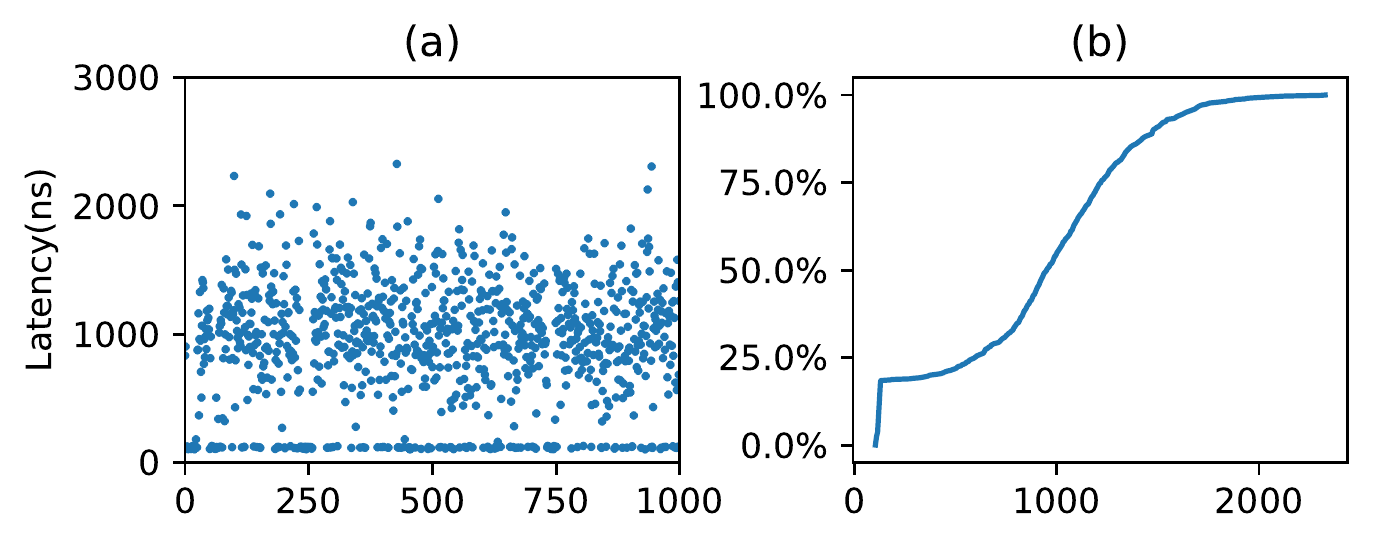}	
	\caption{Sequential Write Latency in APP direct mode.}
	\label{appdirect_rand_write_latency}
\end{figure}

Much different with read latency, We will shows the results of write latency experiments in this section. As introduced in Section 3.2, our benchmarks make all memory write queues full in every level of memory hierarchy, which makes the measured time of write instructions committing equals to the time of one queue entry being issued which nearly equals to memory write latency. Figure \ref{appdirect_seq_write_latency} shows the latency values of sequential writes on the timeline. We can see that, about 10\% writes show low latency at about 170ns. And most write latency values distribute from 200 to 1100ns. We can find a similar pattern in random write experiment, whose result is shown in Figure \ref{appdirect_rand_write_latency}. 

According to our benchmark, we consider the 10\% low values as the latency of write instruction successfully issued once it enters into ROB. To other values, write instructions need to wait for an empty entry in memory queues. Some wait for a long time and some wait shorter. The longest situation is that next deleted write command in memory queue is just issued on AEP module. And in this situation, new write instruction must write for a full AEP writes, whose time is AEP write latency. As a summary, we estimate that, write latency of AEP is nearly 1200ns. 

\subsection{Architecture features of DRAM cache}
In memory mode, we also measure the architecture feature of DRAM cache. We focus on three features: cache tag placement, cache associativity and set index mapping. 

\begin{figure}
	\centering
	\includegraphics[scale=0.55]{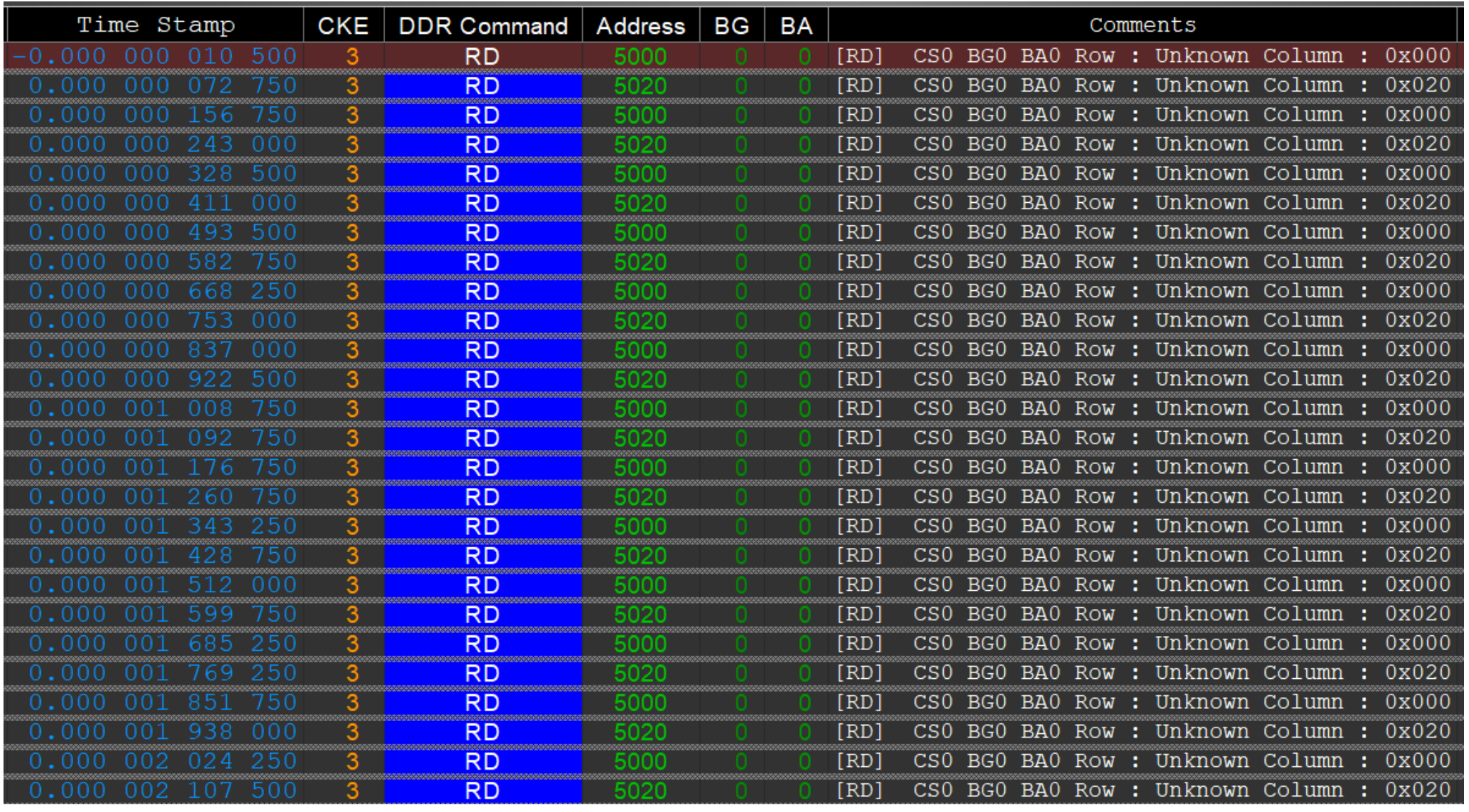}	
	\caption{DDR Commands when DRAM cache hit.}
	\label{lecroy-hit}
\end{figure}

We use Lecroy Kibra DDR Protocol Analyzer Suite to see what happens on DDR bus during a DRAM cache hit or miss. In benchmarks, we use a pair of addresses to test whether they are both cache hit access or both cache misses. In hit case, we catch a fragment of DDR commands as shown in Figure \ref{lecroy-hit}. We can see that all commands are RD (read) and two addresses appear alternately. This means that, a cache hit leads to only one DRAM read. This points to that cache tag and cache data are stored in one cacheline, maybe some bits of ECC are replaced as cache tag.

\begin{figure}
	\centering
	\includegraphics[scale=0.55]{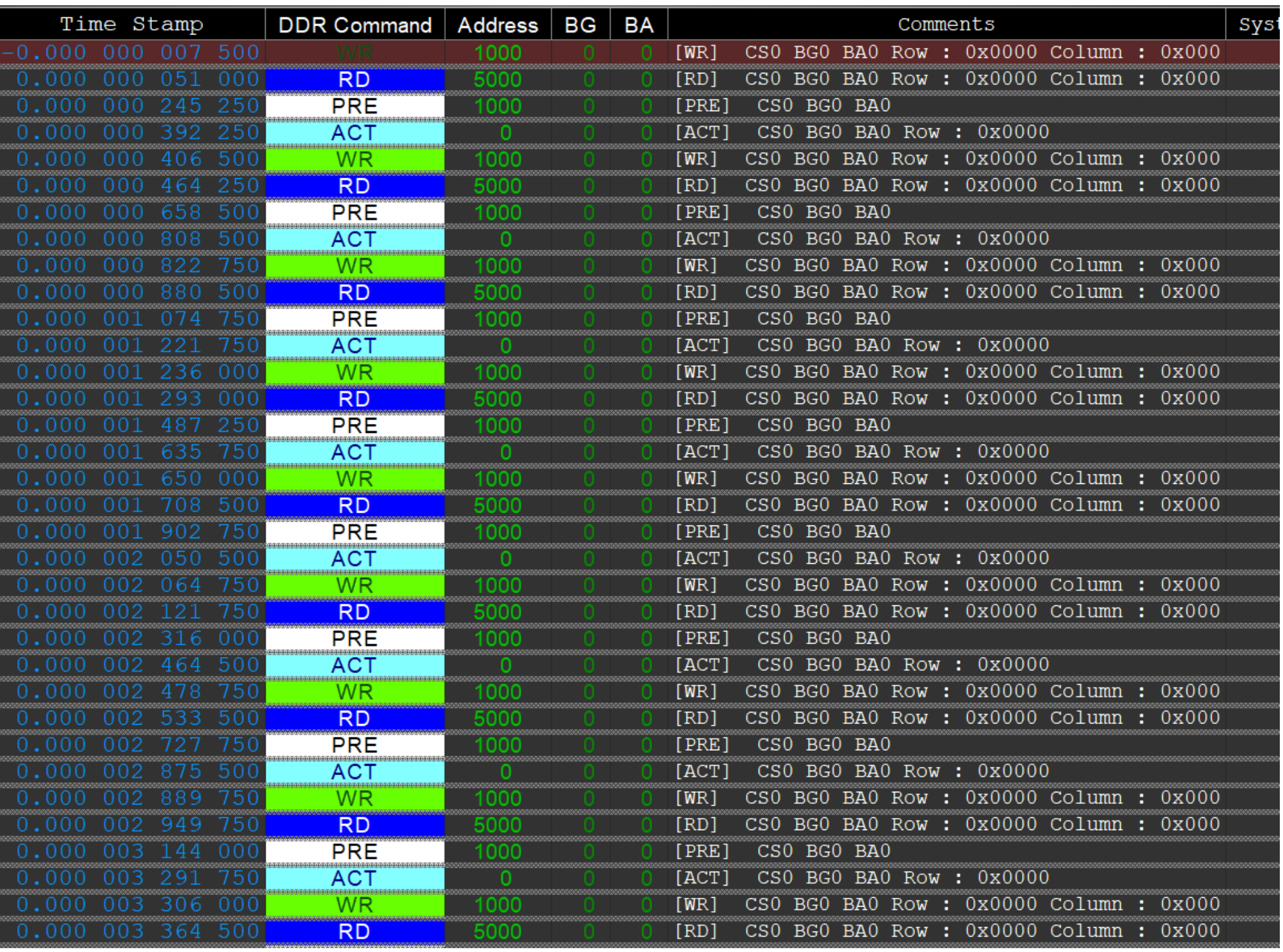}	
	\caption{DDR Commands when DRAM cache miss.}
	\label{lecroy-miss}
\end{figure}

The cache miss case is shown in Figure \ref{lecroy-miss}, where one address corresponds to one read and one write command. In our opinion, read command corresponds to cache tag read. When IMC realizes a cache miss according to cache tag content, it sends a fetch command to AEP and updates cache tag and data in DRAM cache via a DRAM write command.

The conclusion of tag and data are placed in one cacheline leads to another possibility of that DRAM cache is direct-mapped. This is because, all data of one cache set are separate in different cachelines in set-associative cache. So there must be some cases that tag and data are not in one cacheline, but we haven’t observed a phenomenon like this.

\begin{figure}
	\centering
	\includegraphics[scale=0.55]{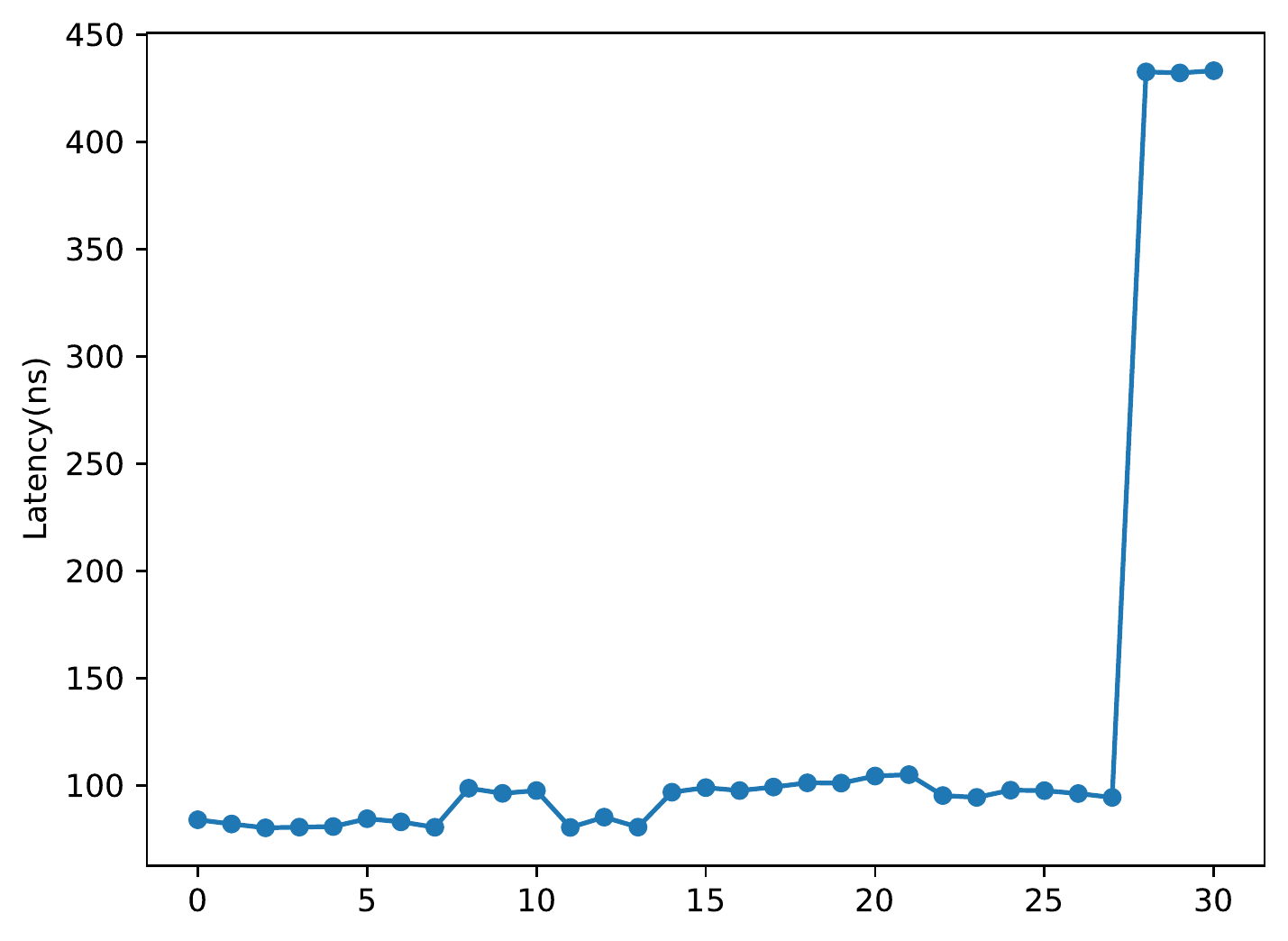}	
	\caption{Measured latency of which bit in memory addresses is different.}
	\label{assoc-latency}
\end{figure}

Another evidence of direct-mapped cache is set index mapping of DRAM cache. As introduced in Section 3.3, we use pair addresses with only one bit difference to test at which time cache is hit or miss. In cache miss case, the measured latency is AEP read latency, or it is DRAM read latency. Figure \ref{assoc-latency} shows the measured latency of which bit in memory addresses is different. 

We can see that when bit of from 0 to 33 changes, the latency stay low, which means all accesses are cache hit. So the set index is simply continuous low address bits. More precisely, as cache granularity is 64B, bit 33-6 is set index as bit 5-0 are cacheline offset. So total number of cache set is 256M, divides to 16GB DRAM capacity, each cache set is 64B capacity. This conclusion also leads to that DRAM cache is direct-mapped.

\section{Summary and future work}
As a summary of our experiments on AEP DIMMs, we find three important features of AEP in this paper which are helpful to guide further AEP optimization:

First, we have found that, rather than a small value, write latency of AEP is about 3 times of read latency in fact, which means write performance of AEP could be bottleneck to some write-intensive programs. Together with bandwidth, AEP has a quite unbalance performance between read and write. Maybe future design optimization should think highly of write optimization. Some prior work had proposed that, part of DRAM cache can be used as write-exclusive buffer, to achieve a higher cache hit rate on write instructions.

Second, DRAM cache is a direct-mapped cache in AEP memory mode. And compared to set-associative cache, which is commonly used on CPU cache, direct-mapped cache performs lower hit rate. On contract, direct-mapped cache has lower cache hit latency when cache tag and data are fetched in one DRAM read, which cannot be achieved easily in set-associative cache. In future works, we should give consideration to both hit rate and hit latency on DRAM cache designs.

At last, a small buffer named XPbuffer benefits both read and write commands on AEP DIMM. But the capacity of XPbuffer is limited by area on DIMM. If we can use XPbuffer more wisely or we can modify DDR protocol to adapt 256B access granularity of AEP, the access efficiency of AEP could grow up further. Some work had proposed that, DRAM cache can be filled and evicted in 256B granularity. This is more suitable to AEP DIMM and it will reduce the storage space of cache tags. 

\section{Conclusion}
In this paper, we propose a new measurement methodology of memory latency and cache architecture. As an application, we measure latency and bandwidth of Intel AEP Optane DIMM and architecture design parameters of DRAM cache with AEP. The release of AEP DIMM has great significance of Non-Volatile Main Memory research but its parameters are not open to most users. According to our evaluation, 
%
%
%
%

\end{document}